\documentclass[a4paper,11pt]{article}
\pdfoutput=1 
\usepackage{jheppub} 
\usepackage{epsfig}
\usepackage{caption}
\usepackage{subcaption}
\usepackage{color,soul,bm}
\usepackage{float}
\usepackage{amsfonts}
\usepackage{url}
\usepackage{slashed}
\usepackage{accents}
\usepackage{bbm,multicol}
\usepackage{lipsum}
\usepackage{mathtools}
\usepackage{physics}
\usepackage{adjustbox}
\graphicspath{{./figs/}}
\usepackage{soul}
\usepackage[normalem]{ulem}

\hypersetup{
  bookmarks=true,         
  unicode=false,          
  pdftoolbar=true,        
 pdfmenubar=true,        
 pdffitwindow=true,     
 pdfstartview={FitH},    
 pdfsubject={Dark Matter},   
 pdfnewwindow=true,      
 pdfcreator={RevTeX},
 colorlinks=true,       
 linkcolor=red,          
 citecolor=blue,        
 filecolor=black,      
 urlcolor=blue,           
  }

\title{Scalaron dark matter dynamics: effects of Higgs non-minimal coupling to gravity}
\author[a]{Shibendu Gupta Choudhury,}
\author[b]{Koushik Dutta,}
\author[c]{and Deep Ghosh}
\affiliation[a]{Centre for Theoretical Physics, Jamia Millia Islamia, New Delhi-110025, India.}
\affiliation[b]{Department of Physical Sciences, Indian Institute of Science Education and Research (IISER) Kolkata, Campus Road, Mohanpur, West Bengal 741246.}
\affiliation[c]{Theory Division, Saha Institute of Nuclear Physics (SINP),
Sector-I, Block-AF, Bidhannagar, Kolkata 700064.}
\emailAdd{pdf.schoudhury@jmi.ac.in}
\emailAdd{koushik@iiserkol.ac.in}
\emailAdd{matrideb1@gmail.com}
\abstract{One of the key features of the $R^2$-gravity is the embedding of a scalar field, \textit{scalaron}, into the gravity sector. The scalaron interacts with the Standard Model (SM) matter fields through Planck-suppressed couplings. If the scalaron serves as a viable dark matter (DM) candidate, it can account for the lack of evidence of DM interactions beyond gravity in experimental and observational probes to date. The realization of the scalaron, as a cold DM candidate, depends on an induced trilinear interaction with the SM Higgs {via its quartic self coupling}. Here, we introduce a Higgs non-minimal coupling to gravity that additionally contributes to the induced trilinear interaction with its existing competing part, originated from the $R^2$-gravity. We study the interplay between these two contributions in the early universe, which determines both the initial conditions and evolution of the scalaron, leading to cold DM behavior at a later epoch. 
The trilinear interaction vanishes at the leading order for certain combinations of the Higgs non-minimal coupling ($\xi$) and the scalaron mass ($m$), thereby setting the scalaron density through \textit{misalignment} mechanism, similar to axions. In this case, the scalaron DM mass is  obtained as, $2.7 ~{\rm meV} \lesssim m \lesssim 0.7 ~\rm{MeV}$. The lower limit on the mass is set by the fifth force constraints from torsion balance experiment, whereas the upper bound arises from INTEGRAL/SPI limits on the excess gamma-ray flux due to possible decay of scalaron into two photons. On the other hand, when the trilinear interaction is non-zero and dominated by the Higgs quartic self coupling, the DM relic density is satisfied with $m \simeq 3.6$ meV. Alternately, when the Higgs non-minimal coupling to gravity dominates, the mass is found to be in the range $10-770$ meV. We also obtain, for the first time within the scalaron-Higgs mixed model, an upper bound on $|\xi m|$ of $1.5\times 10^{17}$ GeV, from Higgs measurements at the LHC.}

\begin{document} 
\maketitle
\flushbottom  
\section{Introduction}
Being in the realm of Einstein's theory of gravitation, a non-baryonic, pressure-less and non-luminous component of matter, termed as dark matter (DM), is needed for large-scale structure (LSS) formation of our universe. Till date, gravitational imprints of DM have been substantiated with a plethora of cosmological and astrophysical phenomena. However, interactions other than gravity are yet to be confirmed from experiments and observations in operation. In particular, DM interactions with other Standard Model (SM) particles are usually probed in direct, indirect and collider searches. However, its non-observation in these experimental facilities indicates \textit{feeble} interaction rate, while challenging the viability of canonical WIMP DM scenario in its simplest possible form \cite{Profumo:2017hqp,Bertone:2004pz}. 

As a nightmare scenario, DM may not interact with SM particles at all, which does not hinder the gravitational clustering, essential for the LSS formation. It only depends on the DM density during the matter-radiation equality epoch, in the cold DM paradigm. Then, the question comes why there is little or absolutely no interaction between DM and SM particles. In this context, extensions of Einstein's general relativity (GR), particularly $R^2$-gravity ($R$ is the Ricci scalar) can provide an explanation to naturally suppressed interactions of DM with the rest of matter fields \cite{Cembranos:2008gj, Corda:2011aa, Katsuragawa:2016yir, Katsuragawa:2017wge, Shtanov:2021uif}. In this case, an extra scalar degree of freedom is hidden inside the gravity sector, in addition to two polarizations of gravitational waves. The scalar field becomes disentangled from the gravity sector, while restoring the usual Einstein-Hilbert action \cite{Nojiri:2017ncd,DeFelice:2010aj}, through a conformal transformation of the original theory with $R^2$ term. Since the scalar field, often termed as \textit{scalaron} \cite{starobinsky,Nojiri:2008nt}, comes from the gravity sector, it couples democratically to all other preexisting matter fields but with Planck-scale suppressed couplings after the transformation. Therefore, the scalaron naturally explains the non-observation of DM-SM interactions in existing experiments till date, subject to its viability as DM candidate.
 
 How can the scalaron density satisfy the observed relic density of DM ? The scalar field dynamics, leading to the present-day DM density is a well studied scenario in the context of axion cosmology \cite{Preskill:1982cy,Abbott:1982af,Dine:1982ah,Marsh:2015xka}. The axion field is \textit{misaligned} initially from its potential minima. When the Hubble friction becomes negligible compared to the axion mass, coherent oscillation  (i.e., associated quanta are in the zero momentum state) starts about the minima of its potential, approximated to a quadratic form. The energy density of the axion field during coherent oscillation redshifts as  pressure-less matter, thereby validating it as a cold DM candidate. The observed DM relic density determines the axion mass, $m \sim \mathcal{O}(\mu \rm eV)$ and the initial field value, $\phi_i \sim 10^{10} {~ \rm GeV}$, for the misalignment scenario during a radiation-dominated universe. 

The misalignment mechanism in the context of scalaron has been discussed in Ref.\cite{Cembranos:2008gj}, without taking effects of scalaron interactions with the SM particles in determining its dynamics in the early universe. An induced trilinear interaction of the scalaron with the Higgs, fixed by the Higgs quartic self-coupling and the vacuum expectation value (VEV), determines the initial condition and the subsequent scalaron dynamics, despite the Planck suppression stemming from its gravitational origin. Thus, the evolution of the scalaron is connected to the Electroweak phase transition (EWPT) \cite{Shtanov:2021uif,Shtanov:2022xew}, during which the Higgs field moves from a trivial minima ($\langle h \rangle=0$) to its SM VEV. In this scenario, the coherent oscillation of the scalaron starts only when the Higgs field acquires its SM VEV, unlike the former case, where the Hubble friction decides the onset of the oscillation. As a result, the scalaron saturates the DM density only for a very restrictive mass around $\mathcal{O}$(meV). In contrast, the scalaron mass for the case in Ref.\cite{Cembranos:2008gj},  varies from $\mathcal{O}(\text{meV})$ to $\mathcal{O}(\text{MeV})$, depending on the initial field value, which is set independently of Higgs dynamics during the EWPT.
 
In the present work, we introduce Higgs non-minimal coupling to gravity ($\xi$) in addition to the $R^2$ term in the Einstein-Hilbert action, for which the above-mentioned scenarios can be realized in a general framework. In particular, the trilinear scalaron-Higgs interaction now gets competing contributions from both $R^2$ term and the Higgs non-minimal coupling to gravity. We find that an interplay between these terms decides the initial condition and the scalaron dynamics in the early universe. When the interaction is dominated by the Higgs quartic coupling, the scalaron mass around a few meV fixes the DM relic density. On the other hand, the scalaron mass can vary from $\mathcal{O}(10-700)$ meV in case of Higgs non-minimal coupling to gravity dominating the interaction. The upper limit on the scalaron mass in this scenario comes from the Higgs mass and signal strength measurements at the Large hadron Collider (LHC).

The trilinear interaction vanishes at the leading order for certain combinations of $\xi$ and the scalaron mass, $m$. Consequently, the scalaron density is set via misalignment dynamics, as in the axion cosmology. In this case, we find that the DM relic density is satisfied for a range of scalaron masses, i.e., from $\mathcal{O}$(meV) nearly to $\mathcal{O}$(MeV). This lower limit on $m$ comes from the fifth force constraint obtained in the precision torsion balance experiment, while the upper bound stems from $16$ years of gamma-ray observations by INTEGRAL/SPI telescope.

The organization of the paper is as follows : the general framework for scalaron dynamics in presence of the Higgs non-minimal coupling is discussed in Sec.\ref{sec:sec2}, followed by the detailed analytic calculation of scalaron dynamics in Sec.\ref{sec:sec3}. In Sec.~\ref{sec:sec4}, we discuss the relevant experimental constraints on the scenarios considered, in particular, the LHC constraints on $|\xi m|$ in our scalaron-Higgs mixed model, followed by the gamma ray constraints on possible decay of the scalaron into a pair of photons from the INTEGRAL/SPI telescope and the fifth force constraint on the scalaron mass. Finally, we summarize our findings in Sec.~\ref{sec:sec5} and present our concluding remarks in Sec.~\ref{sec:sec6}.

\section{The general framework}
\label{sec:sec2}
We first write the Einstein-Hilbert action with an additional $R^2$ term in the Jordan frame as the following \cite{starobinsky, Shtanov:2021uif,Oikonomou:2025htz}:
\begin{equation} 
S_g = - \frac{M_P^2}{3} \int d^4 x \sqrt{-g} \left( R - \frac{R^2}{6 m^2} \right) \, ,
\label{eq:Sg}
\end{equation}
 where $M_P (= \sqrt{3 / 16 \pi G} \approx 3 \times 10^{18}$ GeV) is the normalized Planck mass.\footnote{We have followed the notation of \cite{Shtanov:2021uif} for the normalized Planck mass $M_P$ so that the conformal factor comes out to be $e^{-\frac{\phi}{M_P}}$, instead of $e^{-\sqrt{\frac{2}{3}}\frac{\phi}{M_P}}$.} We do not include the cosmological constant in the gravity lagrangian, since our present study is related to a standard radiation-dominated universe only. At this point, we note that the scale $m$ is going to determine the mass of the scalar field responsible for the DM, and it is in the range of  meV to $\rm MeV$.
 
Now, we assume that the preexisting matter component follows the SM of particle physics and all SM particles other than Higgs, minimally couple to gravity. Higgs non-minimal coupling to gravity has been extensively discussed in the context of Higgs-inflation \cite{Bezrukov:2007ep, Giudice:2010ka,Rubio:2018ogq}. We only write the Higgs part of the matter sector with its non-minimal coupling to gravity as, 
\begin{equation} 
S_M \supset \int d^4 x \sqrt{-g} \left[ g^{\mu\nu} \left( D_\mu \Phi \right)^\dagger D_\nu \Phi - \lambda \left( \Phi^\dagger \Phi - \frac{v^2}{2} \right)^2 -\xi R \left( \Phi^\dagger \Phi - \frac{v^2}{2} \right)\right] \, ,
\label{eq:matter}
\end{equation}
where $\Phi$ is the Higgs doublet, $v ~(= 246~ {\rm GeV})$ is the Higgs VEV at zero temperature and $\lambda$ ($\approx 0.13$ within SM) is the Higgs quartic self-coupling.  We shall demonstrate that the current theory with Eq.~\eqref{eq:Sg} and Eq.~\eqref{eq:matter} will generate an interaction between the scalaron and the Higgs at the renormalizable level. In the context of inflation, such theories with both $R^2$-gravity and Higgs non-minimal coupling terms have been considered in the literature \cite{Ema:2017rqn, Gorbunov:2018llf, He:2018mgb, He:2018gyf,Gundhi:2018wyz, Cheong:2021vdb,Canko:2019mud, Wang:2017fuy, Salvio:2015kka, Salvio:2016vxi}. 

To note, the non-minimal coupling, $\xi$ is a dimensionless quantity, which couples Higgs to gravity at the renormalizable level, thereby potentially can modify the Higgs mass and its decay widths. This coupling solely can not be constrained from the Higgs measurements at LHC within the scalaron-Higgs mixed model, rather we obtain an upper bound on $|\xi m |$ to be $1.5\times 10^{17}$ GeV at $95\%$ confidence level (C.L.). For the detailed derivation of this result see Sec.\ref{sec:sec4lhc}. The sign of $\xi$ can not be determined uniquely from the signal strength, since only the magnitude of $\xi$ appears in the amplitudes of physical processes, measured at the LHC. However, to ensure the vacuum stability of Higgs potential during high scale inflationary epoch, $\xi$ turns out to be a positive quantity, particularly $\xi \gtrsim 0.06$ \cite{Herranen:2014cua}. Also for post-inflationary epoch $\xi$ is reported to be positive \cite{Figueroa:2017slm}. Although we are not concerned with inflationary dynamics in the present context, we work with positive values of $\xi$, i.e., $0\leq \xi m \leq 1.5\times 10^{17}$ GeV.

To express the total action, (i.e., $S=S_g+S_M$) in the Einstein frame, we first introduce an auxiliary field, $\sigma$ and write an equivalent action (well known technique in $f(R)$ gravity theories, see for example Refs.\cite{Faraoni:2004pi, DeFelice:2010aj}) as follows :
\begin{align}\label{eq:aux}
    S= -\frac{M^2_P}{3}\int d^4 x \sqrt{-g} \left[ \left(1-\frac{\sigma}{3m^2}+\frac{3\xi}{M^2_P}\left( \Phi^\dagger \Phi - \frac{v^2}{2} \right)\right)R + \frac{\sigma^2}{6m^2}\right] \\ \nonumber
    + \int d^4 x \sqrt{-g} \left[ g^{\mu\nu} \left( D_\mu \Phi \right)^\dagger D_\nu \Phi - \lambda \left( \Phi^\dagger \Phi - \frac{v^2}{2} \right)^2\right].
    \end{align}
The putative scalar degree of freedom ($\phi$) becomes disentangled from the gravity sector, restoring usual Einstein-Hilbert term via the following conformal transformation \cite{Ema:2023dxm} : 
\begin{equation}
g_{\mu\nu} (x^\mu) = \Omega^{-2} \tilde{g}_{\mu\nu} (x^\mu), \qquad \text{with} \quad \Omega = e^{\phi / 2 M_P}=\left(1-\frac{\sigma}{3m^2}+\frac{3\xi}{M^2_P}\left( \Phi^\dagger \Phi - \frac{v^2}{2} \right)\right)^{1/2}.
\label{eq:conf}
\end{equation}
With this transformation the total action in the Einstein frame becomes,
\begin{align} 
S = &- \frac{M_P^2}{3} \int d^4 x \sqrt{-\tilde{g}} ~\tilde{R} + \int d^4 x \sqrt{-\tilde{g}} \left[ \frac12 \tilde{g}^{\mu\nu} \partial_\mu \phi \partial_\nu \phi - V(\phi,\tilde{\Phi}) \right] \nonumber\\ &+\int d^4 x \sqrt{-\tilde{g}} \, \tilde{g}^{\mu\nu} \left[ \left( D_\mu \tilde{\Phi} \right)^\dagger D_\nu \tilde{\Phi} + \frac{1}{2 M_P} \partial_\mu \left( \tilde{\Phi}^\dagger \tilde{\Phi} \right) \partial_\nu \phi + \frac{1}{4 M_P^2} \tilde{\Phi}^\dagger \tilde{\Phi} \, \partial_\mu \phi \partial_\nu \phi \right]. 
\label{eq:Shn}
\end{align}
To note, the Higgs field is also transformed as $\Phi \to \Omega \tilde{\Phi}$, to keep the kinetic term in its canonical form as in the Jordan frame. The dynamics of scalaron field, $\phi$ is determined by the potential, $V(\phi,\tilde{\Phi})$, which reads as,  
\begin{equation}
 V(\phi, \tilde{\Phi})= \frac12 m^2 M_P^2 \left[1 - e^{- \phi / M_P} -\frac{3 \xi}{M_P^2} \left( \tilde{\Phi}^\dagger\tilde{\Phi} - e^{- \phi / M_P} \frac{v^2}{2} \right)\right]^2+\lambda \left( \tilde{\Phi}^\dagger \tilde{\Phi} - e^{- \phi / M_P} \frac{v^2}{2} \right)^2, 
 \label{potPhi}
\end{equation}
while dropping non-renormalizable terms suppressed by the Planck mass. For sub-Planckian value of $\phi$, i.e., $\phi/M_P \ll 1$ and taking the Higgs doublet in the unitary gauge as $\tilde{\Phi}^T=(0 ~~ h/\sqrt{2})$, the potential with dominant contributions becomes,
\begin{equation} \label{veff}
V(\phi, h) \simeq \frac{1}{2} m^2 \phi^2 + \frac{\lambda}{4} \left( h^2 - v^2 \right)^2 + \frac{(h^2-v^2)~\phi}{2 M_P} \left(\lambda v^2 - 3 \xi m^2\right).
\end{equation}
To note, the trilinear coupling between the Higgs and the scalaron, induced after the transformation, receives two distinct contributions - one from $R^2$ term and other from the Higgs non-minimal coupling. The interplay between these two terms, as we shall see, will determine the scalaron dynamics in the early universe. 

We also incorporate finite temperature correction to the Higgs potential in our scalaron-Higgs effective  theory, given as
\begin{align} \label{Leff}
\mathcal{L}_\text{eff} \supset \frac12 (\partial_\mu \phi)(\partial^\mu \phi)  + \frac12 (\partial_\mu h)(\partial^\mu h) - V(\phi,h) - \frac{1}{6} T^2 h^2,
\end{align}
where the last term depending on the SM bath temperature ($T$) denotes a finite temperature contribution to the Higgs potential, as approximated in Ref.~\cite{Rubakov:2017xzr}.  Within this framework, the scalaron dynamics is studied in conjunction with the Higgs during the EWPT, which sets the initial condition for the scalaron \cite{Shtanov:2021uif}. Therefore, we first 
 find the extrema of the potential via calculating $dV/d\phi =0$ and $dV/d h=0$, which yields,
\begin{align}
&\phi_0 = \frac{h^2-v^2}{2 m^2~ M_P }\left(3\xi m^2 - \lambda v^2\right),\\
& h_0=0 ~\text{or} ~ h_0^2  = v^2 + \frac{\phi }{\lambda M_P} \left(3\xi m^2 - \lambda v^2\right)- \frac{T^2}{3\lambda}.
\end{align}
The non-trivial Higgs minima is modified owing to  scalaron-Higgs interaction. For sub-Planckian values of $\phi$, the Higgs minima can well be approximated to its SM value, evident from the following :
\begin{align}
h_0^2 = v^2 \left(1- \frac{T^2}{T^2_c}\right),~~ T_c=  \sqrt{3 \lambda} v \left[1-\frac{\left(3\xi m^2 - \lambda v^2\right)^2}{2 m^2 M^2_P \lambda}\right]^{1/2} \approx \sqrt{3 \lambda} v,
\label{eq:hmin}
\end{align}
where $T_c$ is the critical temperature of the EWPT within the SM. For $T>T_c$, the minima of the Higgs field is at $h_0=0$ and for $T<T_c$, the minima evolves over temperature and finally settles down to its zero temperature VEV. Consequently, the scalaron minima is dictated by the Higgs dynamics, as manifest in the following,
\begin{align}
\phi_0 (T) =  
\begin{cases}
  \frac{v^2}{2m^2 M_P}\left(\lambda v^2 - 3\xi m^2\right)~~ \text{for $T>T_c$},\vspace{0.2cm}\\
  \frac{v^2}{2m^2 M_P}\left(\frac{T}{T_c}\right)^2 \left(\lambda v^2 - 3\xi m^2\right)  ~~ \text{for $T \leq T_c$}\vspace{0.2cm},\\
  ~0 ~~ \text{for}~ 3\xi m^2 = \lambda v^2 ~ \text{and for all $T$}. 
\end{cases}
\label{eq:min}
\end{align}
It is apparent from the above conditions that for non-zero trilinear interaction with the Higgs, the minima of the scalaron becomes non-trivial before the EWPT ; subsequently the minima shifts towards zero as $T \ll T_c$. Hence, the initial minima of the scalaron  determines the energy density of the coherent oscillation of the field around its final minima, i.e., $\phi_0=0$. This is a quite  departure from the standard misalignment scenario, where the initial condition is usually set away from the minima, which provides the energy of the coherent oscillation. In the present situation, the source of that energy comes from the interaction with the Higgs via $\lambda$ and $\xi$. Thus the scalaron, being in its natural minima at the early epoch, can account for the DM density at a later epoch.  

 In presence of Higgs non-minimal coupling to gravity the trilinear interaction vanish at the leading order for
 $3\xi m^2 = \lambda v^2$. Consequently, we can approximate the scalaron evolution to be decoupled from the Higgs sector from the very beginning and the scalaron minima to be $\phi\approx 0$ for all temperatures. In passing, we note that the interaction does not identically vanish under the present condition, rather leaves a residual term, suppressed by $1/M^3_P$. However, it leads to a small initial field value, $\frac{\phi_0(T_c)}{M_P}=\frac{\lambda^2 v^8}{4m^4 M_p^4}\sim 10^{-12}$, thereby effectively needing an extra initial condition for the scalaron to be a valid DM candidate, as shown in the subsequent study. 
 
 Before studying these two scenarios, we write the equation of motion (EOM) for both the scalaron and Higgs field as follows:\footnote{We are interested in dynamics of zero momentum modes of the scalaron and the Higgs, thereby writing only the EOMs for the homogeneous part.}
\begin{align}
&\ddot \phi + 3 H \dot \phi + m^2 \left[\phi +\frac{h^2-v^2}{2 m^2 M_P}\left(\lambda v^2-3\xi m^2\right)\right] = 0,& \label{eqsf} \\
&\ddot h + 3 H \dot h + \lambda h \left[ h^2 - v^2 \left( 1-\frac{T^2}{T^2_c}\right)+\frac{\phi}{\lambda M_P} \left(\lambda v^2-3\xi m^2\right) \right] = 0,& \label{eqh}
\end{align}    
where the Hubble parameter, \( H = \dot a / a \), in which \( a \) denotes the scale factor. The dynamics of the Higgs field becomes independent from $\phi$ for $\phi \ll M_P$, evident from  Eq.\ref{eqh}. For $T<T_c$, the Higgs field makes a transition from $h=0$ to $h=v$, thus its mass becomes, $m_H \sim \sqrt{\lambda} v$. Therefore, the characteristic time scale of Higgs oscillation is approximately, $t_h \sim  (\lambda v^2)^{-1/2}$, whereas the scalaron counterpart is  $t_s \sim m^{-1}$. For $t_h \ll t_s$, we can approximate the instantaneous value of the Higgs field with its temperature-dependent VEV ($h_0$), as shown in Eq.\ref{eq:hmin} in solving the scalaron field equation. In other words, the oscillation of scalaron field does not affect the Higgs dynamics for $m \ll  m_H$ \cite{Shtanov:2021uif}. Also note that, the scalaron is expected to be a DM candidate, its decay to a pair of Higgs particles is conveniently prevented with this condition. Therefore, the EOM of $\phi$ can be solved independently after replacing the Higgs field with its temperature-dependent VEV into  Eq.\ref{eqsf} which now takes the following form :
\begin{align}
\ddot \phi + 3 H \dot \phi + m^2 \left[\phi - \frac{T^2}{T^2_c}~\frac{\left(\lambda v^2-3\xi m^2\right)~v^2}{2 m^2 M_P}\right] = 0.
\end{align}
As we solve the equation in the radiation-dominated universe from the epoch of the EWPT ($T=T_c$), the time derivatives are conveniently transformed into temperature derivatives, using the relation, $\dot{T}=-HT$, where $H=(\frac{\pi^2}{60}g_*)^{1/2}\frac{T^2}{M_P}$. Here, we neglect the variation of relativistic degrees of freedom ($g_*$). We write the EOM in terms of rescaled dimensionless variables, $\tilde{\phi}=\frac{\phi}{\phi_0 (T_c)}$ (for $\phi_0\neq 0$) and $\tilde{T}=\frac{T}{T_c}$, i.e.,
\begin{align}
\tilde{\phi}'' + \omega^2(\tilde{T}) \left( \tilde{\phi} - \tilde{T}^2 \right) = 0,
\label{eq:shatnov}
\end{align}
where $\phi_0 (T_c)=\frac{v^2}{2m^2 M_P}\left(\lambda v^2 - 3\xi m^2\right)$ and $\displaystyle{\omega^2(\tilde{T})=\frac{m^2}{H(\tilde{T})^2 \tilde{T}^2}}$. In case of $ 3\xi m^2 = \lambda v^2$, the EOM simplifies to 
\begin{align}
\displaystyle{   \bar{\phi}'' + \omega^2(\bar{T})  \bar{\phi} = 0},
    \label{eq:cem}
\end{align}
but with an arbitrary initial condition  away from its minima, i.e., $\phi = \phi_i$ at $T=T_i$, for which we define $\bar{\phi}=\phi/\phi_i$ and $\bar{T}=T/T_i$. Notably, the initial temperature in this scenario can be arbitrary; whereas in the former one, the initial temperature is the critical temperature of the EWPT. This is due to apparent decoupling of the scalaron from the Higgs sector, leaving scalaron dynamics unaffected by the phase transition. 
\section{Scalaron dynamics : effects of Higgs interactions}
\label{sec:sec3}
With the general setup explained earlier, we now proceed to discuss two scenarios: one with a non-zero scalaron–Higgs interaction and the other with a vanishing scalaron–Higgs interaction. We solve  Eq.\ref{eq:shatnov} and Eq.\ref{eq:cem} to demonstrate different dynamics at play, leading the scalaron to be a viable DM candidate. Consequently, the scalaron mass required to saturate the observed dark matter relic density differs widely, depending on the underlying dynamics.
\subsection{Scenario I : $3\xi m^2 \neq \lambda v^2$}
In this scenario, the scalaron stays at its natural minima (as in Eq.\ref{eq:min}) before the EWPT and only starts moving as $T \leq T_c$. The scalaron, initially coupled to the SM thermal bath via the trilinear coupling with Higgs, eventually decouples. However, we will see that the scalaron does not acquire the usual thermal density, rather its initial density is fixed by its initial field value. Subsequently it oscillates \textit{coherently} as a matter field at temperatures well below $T_c$. This is a sheer contrast to axion misalignment scenario, in which the oscillation takes place depending only on the ratio of the Hubble parameter and the axion mass. Here, the onset of the oscillation is governed by the scalaron-Higgs interaction in addition to the Hubble to mass ratio. We demonstrate salient features of scalaron dynamics by solving Eq.\ref{eq:shatnov} semi-analytically with another change of variable, i.e., $ \chi = \tilde{\phi} - \tilde{T}^2$, which yields,
\begin{align}
    \chi'' + \omega^2 (\tilde{T})\left( \chi +\frac{2}{\omega^2(\tilde{T})}\right)\,  = 0.
\end{align}
The equation can be solved analytically in the regime $\tilde{T}\leq 1$, for a constant $g_* (\approx 100)$, which sets $\displaystyle
{\omega^2(\tilde{T}) = \frac{3 m^2 M_P^2}{5 \pi^2 T_c^4 \tilde{T}^6}}$. Initial conditions are set as :  $\chi(\tilde{T}=1)=0$, $\chi'(\tilde{T}=1)=-2$, for $\tilde{\phi}(\tilde{T}=1)=1$ and  $\tilde{\phi'}(\tilde{T}=1)=0$. The above equation further simplifies to a homogeneous one, as $\omega^2 \gg 2$ for $\tilde{T}\leq 1$. Hence, the solution of $\chi$ is given by,  
\begin{align}
\chi(\tilde{T}) &= A \sqrt{\tilde{T}} \left[ J_{\frac{1}{4}}\left(\frac{\sqrt{\alpha}}{2}\right) J_{-\frac{1}{4}}\left( \frac{\sqrt{\alpha}}{2 \tilde{T}^2} \right) - J_{-\frac{1}{4}}\left( \frac{\sqrt{\alpha}}{2} \right) J_{\frac{1}{4}}\left( \frac{\sqrt{\alpha}}{2 \tilde{T}^2} \right) \right],\\
A &= \frac{4}{\sqrt{\alpha}} \Bigg[ J_{-\frac{5}{4}}\left(\frac{\sqrt{\alpha}}{2}\right) J_{\frac{1}{4}}\left(\frac{\sqrt{\alpha}}{2}\right)+ J_{-\frac{1}{4}}\left(\frac{\sqrt{\alpha}}{2}\right) J_{\frac{5}{4}}\left(\frac{\sqrt{\alpha}}{2}\right) \nonumber\\
&\quad  -J_{\frac{1}{4}}\left(\frac{\sqrt{\alpha}}{2}\right) J_{\frac{3}{4}}\left(\frac{\sqrt{\alpha}}{2}\right) - J_{-\frac{1}{4}}\left(\frac{\sqrt{\alpha}}{2}\right) J_{-\frac{3}{4}}\left(\frac{\sqrt{\alpha}}{2}\right) \Bigg]^{-1},
\label{eq:chi}
\end{align}
where \( J_n(x) \) denotes the Bessel function of the first kind of order \( n \), and $\displaystyle{\alpha = \frac{3 m^2 M_P^2}{5 \pi^2 T_c^4}}$.
\begin{figure}[h]
\centering
\includegraphics[scale=1.2]{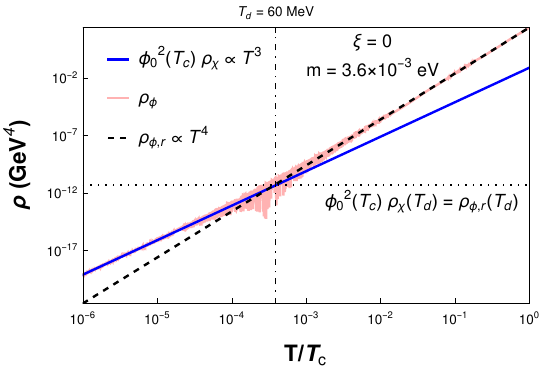}
\caption{ \textit{Scenario-I}: Evolution of the scalaron energy density (shown by the pink solid line) in the absence of non-minimal Higgs coupling to gravity. It first redshifts as radiation density (shown by the black dashed line), then as matter density (shown by the blue solid line). This feature is retained even with non-zero non-minimal Higgs coupling, as long as $3\xi m^2 \neq \lambda v^2$. See text for details.}
\label{fig:fig1}
\end{figure} 
Although the solution of the original field, $\tilde{\phi}$ can be written using the above expression, the solution of $\chi$ gives important insights about the scalaron as a valid DM candidate. For $\tilde{T}<1$, the solution of $\chi$ can be approximated as, $\chi \sim A \tilde{T}^{3/2} \cos\left({\frac{\sqrt{\alpha}}{2\tilde{T}^2}}-\frac{3\pi}{8}\right)$, which in turn sets its (scaled) energy density as the following:
\begin{align}
    \rho_\chi = \frac{1}{2} \dot{\chi}^2 + \frac{1}{2} m^2 \chi^2 \sim \text{const} \cdot~\tilde{T}^3 \propto \left(\frac{a}{a_c}\right)^{-3}.
\end{align}
Here $a_c$ is the scale factor corresponding to $T_c$. We note that the energy density of $\chi$ redshifts as matter throughout the cosmic history, as shown in Fig.\ref{fig:fig1} (blue solid line). The scalaron can be a viable DM candidate if its density follows similar trend as of $\rho_\chi$. We show the energy density of the original field, $\phi$ (depicted by the pink line) in Fig.\ref{fig:fig1}. Evidently, it does not follow $\rho_\chi$ during initial epochs, instead it follows a radiation-like behavior, apparent from the scaling nature shown by the black-dashed line. Such a behavior can be explained from the solution of $\tilde{\phi}$, given by
\begin{align}
    \tilde{\phi}(\tilde{T}) = \tilde{T}^2 + A \sqrt{\tilde{T}} \left[ J_{\frac{1}{4}}\left(\frac{\sqrt{\alpha}}{2}\right) J_{-\frac{1}{4}}\left( \frac{\sqrt{\alpha}}{2 \tilde{T}^2} \right) - J_{-\frac{1}{4}}\left( \frac{\sqrt{\alpha}}{2} \right) J_{\frac{1}{4}}\left( \frac{\sqrt{\alpha}}{2 \tilde{T}^2} \right) \right].
    \label{eq:phi}
\end{align}
 Noticeably, around $T\lesssim T_c$, 
 $\tilde{\phi}(\tilde{T}) \sim \tilde{T}^2$, in turn $\rho_\phi \sim T^4$, while the second term in the equation being subdominant. This is a consequence of the scalaron-Higgs  interaction, however small, keeps the scalaron coupled to the SM radiation bath for a significant duration of its evolution, depending on the scalaron mass in case of $\xi=0$. For $T \ll T_c$, eventually the second term dominates, consequently  the radiation-to-matter transition of the scalaron density takes place at temperature around $T_d=60$ MeV for $m=3.6$ meV (obtained in Eq.\ref{mass} to saturate present day DM density, discussed later), as shown in Fig.\ref{fig:fig1}. This also marks the decoupling of the scalaron from the Higgs, as the Higgs potential takes its usual $T=0$ form. Until the decoupling, the local minima of the scalaron shifts continuously and the scalaron tracks its time-dependent minima, with a characteristic oscillatory feature, apparent from Eq.\ref{eq:phi} and Eq.\ref{eq:min}. This is further depicted in Fig.\ref{fig:fig11}, by showing the evolution of $\phi$ (pink solid line) and its ambient minima (blue dashed line) at earlier epochs before the scalaron decoupling. We emphasize that the oscillatory feature shown in Fig.\ref{fig:fig11} is inherent to scalaron dynamics since the initial epoch, unlike the standard axion scenarios. In passing, we also note that these features of scalaron is retained in presence of non-zero allowed values of $\xi$, as long as the trilinear interaction with the Higgs is non-trivial.

Since, the scalaron energy density at late times, i.e., $T \ll T_c$ matches with that of $\chi$, we can express the scalaron relic density in terms of the (scaled) energy density of $\chi$ at $T=T_c$ as
\begin{figure}
    \centering
    \includegraphics[height=7 cm, width =10 cm]{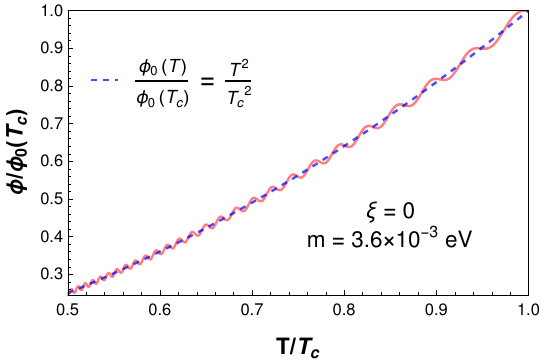}
    \caption{ \textit{Scenario-I}: Evolutions of the scalaron field (pink solid line) and its minima (blue dashed line) at early epochs. To note, the scalaron follows its minima that shifts continuously  due to induced trilinear interaction with the SM Higgs. See text for details.}
    \label{fig:fig11}
\end{figure} 
\begin{align}
    \Omega_\phi = \frac{\rho_\phi (T_0)}{\rho_{c,0}} = \frac{\phi_0^2(T_c)\rho_\chi (T_0)}{\rho_{c,0}}= \frac{\phi_0^2 (T_c)~\rho_\chi (T_c)}{\rho_{c,0}}\left(\frac{T_0}{T_c}\right)^3 \frac{g_{*S} (T_0)}{g_{*S} (T_c)},
\end{align}
where $\rho_{c,0} (\approx 8.2 \times 10^{-47} h^{-2} {~\rm GeV^4} )$ and $T_0$ are the critical energy density and the photon temperature of the present universe, while $g_{*S}(T_i)$ corresponds to the effective number of degrees of freedom contributing to the entropy density of the universe at $T=T_i$. In passing, we note that the decoupling temperature of the scalaron does not explicitly enter into the above expression; however, the decoupling must happen before the matter-radiation equality epoch, i.e., $T \sim 1$ eV, necessary condition for successful structure formation. In the present context, the decoupling happens around $60$ MeV (with $\xi=0$ and $m=3.6$ meV), for which the scalaron density trivially satisfies the Big Bang Nucleosynthesis bound \cite{Fields:2019pfx} on any extra radiation density present in the cosmic soup. 

In calculating the relic density, the initial kinetic energy of $\chi$ only contributes, since at $T=T_c$, $\chi=0$, which renders $\rho_\chi(T_c) =\dot{\chi}^2(T_c)/2=2 H^2_c \approx 2.05\times 10^{-27} ~\rm {GeV^2}$. In addition, we need the initial field value, $\phi(T_c)$, which is determined by both $\lambda$ and $\xi$. Depending on their relative contribution, there are two distinct subclasses within scenario-I, namely $\xi \ll \frac{\lambda v^2}{3 m^2} $ and $\xi \gg \frac{\lambda v^2}{3 m^2}$.

For $\xi \ll \frac{\lambda v^2}{3 m^2}$, $\phi_0 (T_c)=\frac{\lambda v^4}{2 m^2 M_P} = 7.9\times 10^{-11} {\rm GeV} \left(\frac{{~\rm GeV}}{m}\right)^2$, and the observed DM relic density \cite{Planck:2018vyg} is set by the scalaron mass only, i.e.,  
\begin{align}\label{mass}
    \Omega_\phi h^2= 0.12 \left( \frac{3.6\times 10^{-12}{~\rm GeV}}{m}\right)^4,
\end{align}
where we have used $g_{*S} =3.91$ and $g_{*S}(T_c)=g_*(T_c)=100$. This reproduces the primary result of Refs.\cite{Shtanov:2021uif,Shtanov:2024nmf,Shtanov:2025nue} with $\xi=0$ within our general treatment of scalaron dynamics. 

 For $\xi \gg \frac{\lambda v^2}{3 m^2}$, the scalaron starts from a negative field value, $\phi_0 (T_c) \approx -3\xi v^2/2M_P$, unlike the previous one. In this case,  the observed DM relic density is saturated by $\xi$ only, as follows :
\begin{align}
   \Omega_\phi h^2 = 0.12 \left(\frac{\xi}{ 1.95\times 10^{26}}\right)^2.
   \label{eq:xi}
\end{align}
Is such a large required value of $\xi$ allowed in the scalaron-Higgs mixed model ? We have discussed the issue in great detail in Sec.\ref{sec:sec4lhc} and its implication on the scalaron parameter space in Sec. \ref{sec:sec5}. We shall see that indeed such a large value of $\xi$ is allowed as long as the scalaron mass is  \textit{sufficiently} small. 

Noticeably, the scalaron dynamics is completely controlled by the  Higgs couplings ($\lambda$ and $\xi$) for $3\xi m^2 \neq \lambda v^2$. In fact, the initial energy density of the scalaron depends on the Hubble parameter at $T=T_c$ and $\phi_0$ which is determined by the non-minimal coupling $\xi$ and the SM value of $\lambda$. The scalaron saturates the DM density for a relatively wide range of masses when the trilinear interaction becomes either \textit{sufficiently} small or vanishes (to a good approximation) for  $\lambda v^2 = 3\xi m^2$, which is studied next.

To note, the scenario-I is studied assuming the EWPT as a second order transition, which is a valid assumption for scalaron mass around a few meV\cite{Shtanov:2021uif,Shtanov:2022xew}.  However, in Ref.\cite{Shtanov:2024nmf}, the author points out how the scalaron dynamics needs to be modified for higher masses with a realistic scenario of the EWPT, which is a smooth cross-over. In addition, effects due to the QCD trace anomaly has also been discussed in Ref.\cite{Shtanov:2024nmf}.
\begin{figure}[h]
\hspace{-0.9cm}
\includegraphics[scale=0.9]{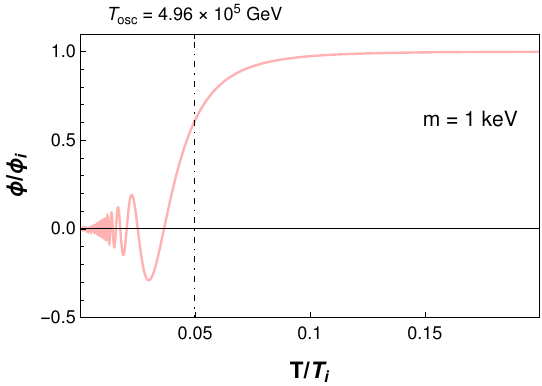}
\includegraphics[scale=0.9]{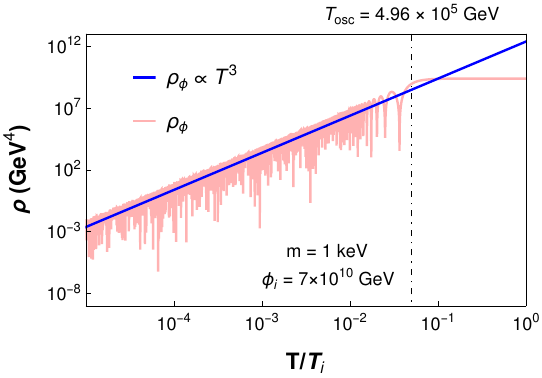}
\caption{\textit{Scenario-II}: Evolution of the scalaron field ($\phi/\phi_i$) (\textit{left panel}) and its energy density ($\rho_{\phi}$) (\textit{right panel}) for $\xi \sim \mathcal{O}(10^{15})$, which fixes $m \sim \mathcal{O}$(keV). The initial field value $\phi_i$, is determined from the observed DM relic density. See text for details.}
\label{fig:fig2}
\end{figure}
\subsection{Scenario II : $3\xi m^2 = \lambda v^2$}
 In this scenario, the trilinear interaction between Higgs and the scalaron can vanish at the leading order for a wide range of allowed values of $\xi$ and $m$, thereby the scalaron remains decoupled from the SM thermal bath at the outset of the evolution. Hence, its dynamics is not controlled by the EWPT, unlike scenario-I. The scalaron density remains trivial if it stays at its minima, $\phi_0=0$. Therefore, the scalaron has to be misaligned from its minima with nearly zero velocity ($\dot{\phi}=0$), which facilitates coherent oscillation of the field around its trivial minima at a later time. This is apparent from the solution of Eq.\ref{eq:cem}, i.e.,
 \begin{align}
 \frac{\phi (\bar{T})}{\phi_i}=  -\frac{\pi  \sqrt{\alpha \bar{T}}} {2 \sqrt{2} }\left[J_{\frac{5}{4}}\left(\frac{\sqrt{\alpha }}{2}\right) J_{-\frac{1}{4}}\left(\frac{\sqrt{\alpha }}{2 \bar{T}^2}\right)+J_{-\frac{5}{4}}\left(\frac{\sqrt{\alpha }}{2}\right) J_{\frac{1}{4}}\left(\frac{\sqrt{\alpha }}{2 \bar{T}^2}\right)\right],
\end{align}
where $\displaystyle{\alpha = \frac{3 m^2 M_P^2}{5 \pi^2 T_i^4}}$ and $\phi_i$ being the (misaligned) field value at the initial epcoh, $T=T_i$. For $\bar{T} < 1$, the solution is approximated again as, $\displaystyle{\phi \sim \phi_i \bar{T}^{3/2} \cos\left({\frac{\sqrt{\alpha}}{2\bar{T}^2}-\frac{3\pi}{8}}\right)}$, which ensures the cold DM behavior of the scalaron density from the onset of the coherent oscillation, unlike the previous scenario, in which there is at first radiation-like behavior, followed by the matter-like one. Here, as long as the Hubble friction is dominant, $\rho_\phi$ remains constant at its initial value, $\rho_{\phi} (T_i) = \frac{1}{2} m^2 \phi^2_i$. The density starts decreasing once the Hubble friction becomes comparable with the scalaron mass and subsequently the scalaron oscillates coherently. The temperature at which coherent oscillation starts is estimated from $3H \sim m$ is given by,
\begin{align}
    T_{osc} \approx \left(\frac{m M_P}{3}\right)^{1/2} \left(\frac{60}{\pi^2 g_*(T_{osc})} \right)^{1/4}.
    \label{eq:osc}
\end{align}
In Fig.\ref{fig:fig2}, evolutions of the scalaron field (left panel) and its density (right panel) have been shown for a typical scalaron mass, $m=1$ keV. It is apparent that the oscillation starts around, $T_{osc} = 4.96 \times 10^5$ GeV, in agreement with Eq.\ref{eq:osc}. As $\rho_\phi$ is approximately constant from $T_i=10^7$ GeV until $T=T_{osc}$, i.e., $\rho_{\phi}(T_{osc})\approx \rho_{\phi}(T_i)$. So, the relic density can be expressed in terms of $\phi_i$ and $m$, utilizing Eq.\ref{eq:osc}, as follows :
\begin{align}
   \Omega_\phi = \frac{\rho_{\phi} (T_i)}{\rho_{c,0}}  \left(\frac{T_0}{T_{osc}}\right)^3 \frac{g_{*S} (T_0)}{g_{*S} (T_{osc})}\implies \Omega_\phi h^2 \approx 0.12 \left(\frac{m}{10^{-6}{~\rm GeV}}\right)^{1/2} \left(\frac{\phi_i}{7 \times 10^{10} {~\rm GeV}}\right)^2,  
\end{align}   
where we have used $g_{*S}(T_{osc})=g_*(T_{osc})=100$. We can now choose $\xi$, which fixes the scalaron mass through the relation, $3 \xi m^2 = \lambda v^2$, since all other parameters are fixed at their SM value. Consequently, the remaining free parameter, $\phi_i$, is determined from the observed DM relic density. In this case, the scalaron mass can vary in a wide range of values with corresponding value of $\phi_i$, unlike the previous scenario, especially for $\xi\ll \lambda v^2/3m^2$. We demonstrate the allowed parameter space of $m$ and $\phi_i$ in Sec.\ref{sec:sec5} and in Fig.\ref{fig:fig3}.
\section{Constraints}
\label{sec:sec4}
Having explained the scalaron dynamics in the previous section, we now elucidate the generic constraints on the scalaron mass $m$ and the Higgs non-minimal coupling to gravity $\xi$, and summarize the results using Fig.\ref{fig:constraint}. These constraints are common to both scalaron dynamics scenarios considered in this work. We provide a detailed discussion on these constraints in the subsequent subsections \ref{sec:sec4lhc}, \ref{sec:sec4ph} and \ref{sec:sec4ff}.
\subsection{LHC Constraints on $\xi$}
\label{sec:sec4lhc}
As already explained, the role of Higgs non-minimal coupling is pivotal in deciding the scalaron dynamics; however the bound on $\xi$ in the non-inflationary scenario is not well studied. In Ref.\cite{Atkins:2012yn} the upper bound on $|\xi|$ derived from the Higgs global signal strength measurement by the LHC has been reported, in the absence of $R^2$-gravity. In the current situation, the scalaron entangles both $R^2$-gravity contribution and the contribution from Higgs non-minimal coupling to gravity, for which an independent bound on $|\xi|$ becomes obscure. However, we find the upper bound on $|\xi m|$ from Higgs measurements, after making appropriate field transformations, details of which  are given below.

 We first write the conformal factor of Eq.\ref{eq:conf} introducing a new field, $\psi=-\frac{\sigma M_p}{3 m^2}$, as the following
\begin{equation}
    \Omega^2=\left[1+\frac{\psi}{M_P}+\frac{3\xi}{M^2_P}\left( \Phi^\dagger \Phi - \frac{v^2}{2} \right)\right],
\end{equation}
where $\psi$ represents the scalar degree of freedom associated with $R^2$-gravity. With this, we can recast the matter part of Eq.\ref{eq:Shn} in terms of $\psi$ and the bare Higgs field as the following.
\begin{align}\label{smnew}
  S_M=\frac{M^2_P}{3}\int \frac{d^4 x \sqrt{-\tilde{g}}}{\Omega^4}\Bigg[ &\Omega^2\tilde{g}^{\mu\nu} \left( D_\mu \Phi \right)^\dagger D_\nu \Phi +\frac{1}{2}\tilde{g}^{\mu\nu}\partial_\mu \psi\partial_\nu \psi+\frac{9\xi^2}{2M_P^2}\tilde{g}^{\mu\nu}\partial_\mu\left(\Phi^\dagger\Phi\right)\partial_\nu\left(\Phi^\dagger\Phi\right)\nonumber\\ &+\frac{3\xi}{M_P}\tilde{g}^{\mu\nu}\partial_\mu \psi\partial_\nu\left(\Phi^\dagger\Phi\right)-\frac12 m^2 \psi^2 -\lambda \left( {\Phi}^\dagger {\Phi} - \frac{v^2}{2} \right)^2\bigg].
\end{align}
We note that a kinetic mixing between $\psi$ and the Higgs emerges at $T\leq T_c$, via the non-minimal Higgs coupling, $\xi$. Our main interest is in the quadratic part of the theory, which is given by, 
\begin{align}
    \mathcal{L}_2=\frac{1}{2}\left(1+{9x^2}\right)\partial_\mu\zeta \partial^\mu\zeta+\frac{1}{2}\partial_\mu\psi \partial^\mu \psi+{3x}\partial_\mu \zeta\partial^\mu \psi - \frac{1}{2}\left(M_\zeta^2\zeta^2+m^2\psi^2 \right) ,
\end{align}
where we have used, $\Phi^T =(0 ~~ (v+\zeta)/\sqrt{2})$, $x = \xi v/M_P $ and $M_\zeta = \sqrt{2 \lambda}v$. Evidently, the redefinition of the kinetic term of the Higgs can not be performed independently as in Refs.\cite{Bezrukov:2007ep,Atkins:2012yn,Boyanovsky:2017esz} due to  the kinetic mixing. Therefore, we make the following basis transformation to get canonical kinetic terms for both the fields. 
\begin{align}
    \eta_1&=\sqrt{1+9x^2}\left(\zeta+\frac{3x}{1+9x^2}\psi\right), \\ \nonumber
    \eta_2&=\frac{\psi}{\sqrt{1+9x^2}}.
\end{align}
Although we get rid of the kinetic mixing, mass mixing between two fields appear in the new basis, as follows :
\begin{equation}
    \mathcal{L}_2=\frac{1}{2}\partial_\mu\eta_1 \partial^\mu\eta_1 +\frac{1}{2}\partial_\mu\eta_2 \partial^\mu\eta_2-\frac12 M_1^2\eta_1^2-\frac{1}{2}M_2^2\eta_2^2-M_{12}^2\eta_1\eta_2,
    \end{equation}
where,
\begin{align}
    M_1^2&=\frac{M_\zeta^2}{1+9x^2},\\ \nonumber
    M_2^2&= \frac{9x^2 M_\zeta^2}{1+9x^2}+m^2\left(1+9x^2\right),\\ \nonumber
    M_{12}^2&=-\frac{3x M_\zeta^2}{1+9x^2}.
\end{align}
Now, we introduce a rotation of the fields to diagonalize the mass matrix, $M_S$, i.e.,
\begin{equation}
    \begin{pmatrix}
        \phi_1 \\ \phi_2
    \end{pmatrix}= \begin{pmatrix} \cos{\theta} & \sin{\theta} \\ -\sin{\theta} & \cos{\theta}\end{pmatrix}\begin{pmatrix}
        \eta_1 \\ \eta_2
    \end{pmatrix},~
    M_S=\begin{pmatrix} M_1 & M_{12} \\ M_{12} & M_2\end{pmatrix},
\end{equation}
where the mixing angle and the mass eigenvalues are given by,
\begin{align}
    \cos{2\theta}=\frac{M_\zeta^2(1-9x^2)-m^2(1+9x^2)^2}{\sqrt{M_\zeta^4(1+9x^2)^2+m^4\left(1+9x^2\right)^4+2m^2M_\zeta^2\left(9x^2-1\right)(1+9x^2)^2}},\nonumber\\ 
     \sin{2\theta}=\frac{-6xM_\zeta^2}{\sqrt{M_\zeta^4(1+9x^2)^2+m^4\left(1+9x^2\right)^4+2m^2M_\zeta^2\left(9x^2-1\right)(1+9x^2)^2}},\nonumber\\
      \mathcal{M}^2_{1,2}=\frac12 \left[M_\zeta^2+m^2\left(1+9x^2\right)\pm \sqrt{M_\zeta^4+m^4\left(1+9x^2\right)^2+2m^2M_\zeta^2\left(9x^2-1\right)}\right].
      \label{eq:mix}
\end{align}
 With these transformations, we arrive at a basis in which the kinetic terms are canonical and the mass matrix is diagonal. We identify $\phi_1$ as the SM Higgs-like field with mass eigenvalue
\begin{equation}
\mathcal{M}_1 = M_\zeta \sqrt{1+\frac{9x^2m^2}{M_\zeta^2}},
\end{equation}
where we have assumed $m\ll M_\zeta$. This assumption is consistent with the scalaron dynamics and, in particular, prevents the decay of the scalaron into two Higgs particles.  Up to this point, we have not imposed any assumption on $x$ and have kept it arbitrary. However, for $\phi_1$ to be the physical Higgs and being consistent with the scalaron dynamics explained in previous sections, $|x m/M_\zeta|$ should be small. This can be readily seen from the expressions of the mixing angle in Eq.~(\ref{eq:mix}) approximated as,
\begin{equation}
    \cos{\theta}\approx\frac{-1}{\sqrt{1+9x^2}}\left(1-\frac{9m^2x^2}{M_\zeta^2}\right),~~~~ \sin{\theta}\approx\frac{3x}{\sqrt{1+9x^2}}\left(1+\frac{m^2}{M_\zeta^2}\right).
\end{equation}
Using these relations, the new fields can be expressed in terms of the original ones as
\begin{align}
\phi_1 &=\sqrt{1+9x^2}\cos{\theta}~\zeta+\frac{3x \cos{\theta}+\sin{\theta}}{\sqrt{1+9x^2}}\psi\approx-\left(1-\frac{9m^2x^2}{M_\zeta^2}\right)~\zeta+\frac{3x m^2}{M_\zeta^2}\psi  \nonumber\\
    \phi_2 &=-\sqrt{1+9x^2}\sin{\theta}~\zeta+\frac{-3x \sin{\theta}+\cos{\theta}}{\sqrt{1+9x^2}}\psi\approx-3x\left(1+\frac{m^2}{M_\zeta^2}\right)~\zeta-\psi.
\end{align}
To note, the cosmology of $\phi_2$ field will be similar to original scalaron field ($\phi$), as defined in Sec.\ref{sec:sec2}, since the energy density is determined by its mass, which is identical to the scalaron mass, i.e., $\mathcal{M}_2 = m$. This is also apparent from the above relations that $\phi_1$ acts just like the SM Higgs and $\phi_2$ as scalaron, having contribution from both Higgs and the additional scalar degree from $R^2$ gravity.

The advantage of this new basis is that we get our SM-like Higgs in its mass (physical) basis with canonical kinetic terms, which is essential to derive relevant bounds from experimental results on new physics scenarios, like the present one. In the present theory, the original Higgs field is just rotated by a factor $\kappa$ (shown in Eq.\ref{eq:kappa}), since $3x m^2/M^2_\zeta \rightarrow 0 $, even with highest possible scalaron mass, i.e., $m=0.7$ MeV, stemming independently from the indirect detection constraints. The effect is captured in the Higgs signal strength measurement at LHC,  which is a standard practice to quantify any possible deviation from the SM prediction. In a narrow-width approximation of Higgs production cross-section and its decay, the Higgs signal strength of a process, $ii\rightarrow h \rightarrow ff$ is given by,
\begin{equation}
    \mu_{if} = \frac{\sigma(ii \rightarrow h) \cdot \mathrm{BR}(h \rightarrow ff)} { \sigma_\mathrm{SM}(ii \rightarrow h) \cdot \mathrm{BR}_\mathrm{SM}(h \rightarrow ff)}.
\end{equation} 
In our scenario, $\mu_{if}=\kappa^2$ and it is given by,
\begin{equation}
    \kappa^2=\left(1+\frac{18m^2x^2}{M_\zeta^2}\right).
    \label{eq:kappa}
\end{equation}

Now, we can use the Higgs mass measurement at the LHC, $\mathcal{M}_1=125.11\pm0.11$ GeV \cite{ATLAS:2023oaq}, obtained from the combined analysis of the decay channels $H\to ZZ^{*} \to 4\ell$ and $H\to\gamma\gamma$, together with the global Higgs signal strength measurements by ATLAS ($\mu =1.05\pm 0.06$ \cite{ATLAS:2022vkf}) and CMS ($\mu= 1.002\pm 0.057$ \cite{CMS:2022dwd}) and find allowed region of $\lambda$ and $|x m|$, which are free parameters of the theory in its current form (see e.g., Refs.~\cite{Costa:2015llh, Robens:2015gla}).  Applying the $95\%$ C.L. allowed ranges for these experimental constraints, we have carried out a parameter scan and checked that $\lambda$ is restricted within a very small region, i.e., $0.117 \lesssim \lambda \lesssim 0.13$ to satisfy the Higgs mass and the global signal strength simultaneously. This combined analysis yields a robust upper bound of $xm \lesssim 12.6$ GeV. This translates to :  $|\xi m| < 1.5\times 10^{17}~\mathrm{GeV}$ within our scalaron-Higgs mixed model. For further details see Appendix.~\ref{app} and Fig.~\ref{fig:placeholder}.  Implications of this bound on scalaron DM scenarios have been discussed in Sec.\ref{sec:sec5} and in Fig.\ref{fig:constraint}. We note that the non-minimal coupling $\xi$ can take very large values, exceeding the bound reported in Ref.~\cite{Atkins:2012yn} by several orders of magnitude, provided the scalaron mass is sufficiently small. Furthermore, the perturbativity of the scalaron-Higgs model (particularly for Higgs quartic self coupling  \cite{Ema:2017rqn}) is also ensured with the current bound, i.e., $\xi^2 m^2/M^2_P \approx 0.002 \ll 4\pi$.
 \begin{figure}[t]
\centering
\includegraphics[width = 10 cm, height = 6 cm]{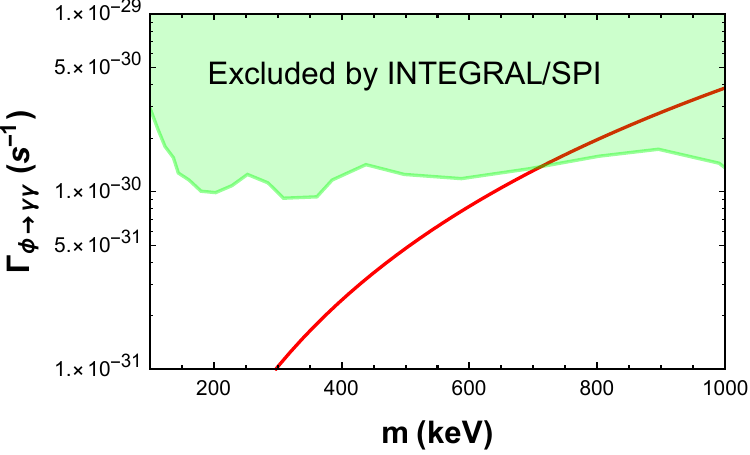}
    \caption{\textit{Constraint on the scalaron decay rate} : The scalaron decay rate (red solid line) as a function of its mass, compared with the exclusion limit at $95\%$ C.L. (green shaded region) on the light DM decaying into a pair of photons, taken from Ref.\cite{Calore:2022pks}. See text for details.}
    \label{fig:decay}
\end{figure}

\subsection{Constraints on the scalaron decay to photons}
\label{sec:sec4ph}
The viability of the scalaron as a DM candidate hinges on its lifetime to be comparable to the age of the universe. Since the scalaron universally couples to the SM matter fields, it can decay to most of the SM particles, if allowed by kinematics. To note, for $m < 2 m_e$, $m_e$ being the electron mass, the only possible decay is to a pair of photons, whose rate should be such that the scalaron becomes cosmologically stable. Although the scalaron does not couple to massless gauge fields at the tree-level due to Weyl invariance,\footnote{The photon action, $\int d^4 x\sqrt{-g} ~F_{\mu\nu}F^{\mu\nu}$ is invariant under transformation defined in Eq.\ref{eq:conf}} the coupling is generated at one loop-level, primarily due to interactions with fermions and massive gauge bosons, as follows \cite{Goldberger:2007zk,Cembranos:2008gj,Shtanov:2025nue,Katsuragawa:2017wge} : 
\begin{equation}
    \mathcal{L}_{\phi-F}=\frac{\phi}{2M_P}\frac{\alpha_{EM}c_{EM}}{8\pi}F_{\mu\nu}F^{\mu\nu},
\end{equation}
where $\alpha_{EM} (=1/137)$ is the electromagnetic fine structure constant and $c_{EM}$ encodes the loop-induced factors. This is reminiscent of the axion–photon coupling, but with a fixed interaction strength stemming from its gravitational origin. Consequently, the decay rate of scalaron to two photons is dependent only on the scalaron mass and given by, 
\begin{equation}       
\Gamma_{\phi\rightarrow \gamma\gamma}=3.8\times 10^{-33} s^{-1}\left(\frac{m}{100 ~{\rm keV}}\right)^3\,,
\label{eq:ph}
\end{equation}
where $c_{EM} = 11/3$ \cite{Cembranos:2008gj}, for scalaron being lighter than the SM fermions and gauge bosons. Evidently, the bound on the lifetime of DM, i.e., $\tau_{DM} \gtrsim 5\times 10^{18} s$ \cite{Audren:2014bca} is easily satisfied by the scalaron. However, the photon flux produced from scalaron decays can constrain the decay rate from the observation of an excess photon flux over known astrophysical backgrounds. In this case, gamma-ray observations with SPI, a spectrometer telescope aboard the INTEGRAL satellite, operating in the energy range of 18 keV to 8 MeV \cite{Winkler:2003nn} will be most sensitive. The photon signal from non-relativistic decays of scalaron in the present epoch yields a monochromatic spectrum, peaking at a single energy bin around, $m/2$. Consequently, the scalaron mass is  constrained from the INTEGRAL/SPI data, unlike scenarios with axion-like particles (ALPs) with an undetermined U(1) symmetry breaking scale involved \cite{Alonso-Alvarez:2019ssa,Arias:2012az}. Nevertheless, there are existing limits on the decay rate of light DM decaying to two photons \cite{Calore:2022pks,Fischer:2022pse}, which can be used to constrain our scalaron scenario. In Fig.\ref{fig:decay}, we show the exclusion region on the decay rate (green shaded region) using the INTEGRAL/SPI data, taken from Ref.\cite{Calore:2022pks}. Authors of Ref.\cite{Calore:2022pks} analyzed 16 years of the INTEGRAL/SPI data, taking the NFW profile of galactic DM density to constrain the decay rate at 95\% C. L. We compare the decay rate of the scalaron (shown in red solid line) as derived in Eq.\ref{eq:ph} to find the upper bound of the scalaron mass as $m = 0.7$ MeV. This bound is derived with the assumption that the scalaron satisfies total DM density, otherwise the exclusion limits become weaker. This also ensures that the scalaron can not decay to a electron-positron pair, because the mass required for that is already excluded from the excess photon flux limit.

 \subsection{Fifth-force constraints}\label{sec:sec4ff}
The scalaron can mediate an attractive fifth force between two non-relativistic particles with masses $m_a$ and $m_b$. The resulting Yukawa-type
potential takes the form \cite{Kapner:2006si, Adelberger:2006dh}
\begin{equation}
V_{ab}(r)
=
-\frac{3\alpha}{16\pi M_P^2}\,
\frac{m_a m_b}{r}\,
e^{-m r},
\end{equation}
where $m$ is the scalaron mass and the dimensionless strength parameter is fixed to
$\alpha = 1/3$ in $R^2$ gravity \cite{Stelle:1977ry}. 

Precision torsion-balance experiments searching for deviations from Newtonian
gravity have not observed any additional Yukawa-type interaction with
gravitational strength ($\alpha \sim \mathcal{O}(1)$) for force ranges exceeding
$38.6~\mu\mathrm{m}$ at $95\%$ C. L., as reported in
Ref.~\cite{Lee:2020zjt}. Interpreting these null results in terms of the above
Yukawa potential with $\alpha = 1/3$ leads to a lower bound on the scalaron mass
\cite{Lee:2020zjt, Kapner:2006si, Adelberger:2006dh, Cembranos:2008gj}
\begin{equation}
    m \gtrsim 2.7~ \mathrm{meV}
    \qquad (95\%~\text{C.L.}) .
\end{equation}
We emphasize that this constraint is specific to $R^2$ gravity, where the scalaron
couples universally to matter with a fixed strength $\alpha = 1/3$.
For a generic ultralight scalar field, such as an axion-like particle, the Yukawa
coupling strength $\alpha$ is a free parameter and can be arbitrarily small.
In such cases, fifth-force constraints gets significantly weakened or entirely
evaded by suppressing the coupling to matter \cite{Calmet:2020rpx, Klimchitskaya:2020cnr, Hees:2018fpg}.

\begin{figure}[htb!]
    \centering
\includegraphics[scale=0.6]{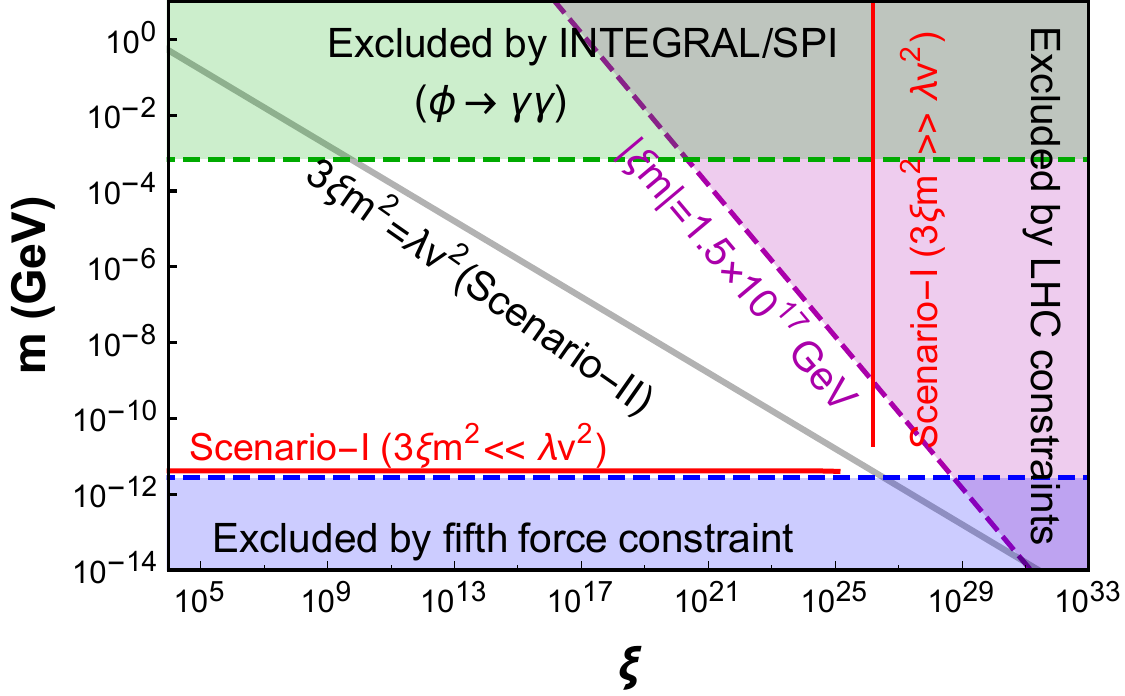}
   \caption{ \textit{Summary of constraints on $m$ and $\xi$}: The relic density constraint in scenario-I (non-zero scalaron-Higgs interaction) is shown by the red solid lines, whereas that in scenario-II (negligible scalaron-Higgs interaction) is shown by the thin gray patch. Constraints from the LHC, the torsion balance experiment and the INTEGRAL/SPI telescope are shown using purple dashed, blue dashed and green dashed lines respectively. These constraints allow scalaron mass ranging from $2.7$ meV to $0.7$ MeV and non-minimal coupling $0\leq\xi \lesssim 10^{26}$.}
    \label{fig:constraint}
\end{figure}
\section{Summary of results}
\label{sec:sec5}
As we have seen in the previous section, there are three independent bounds that restrict the allowed region in the $m$--$\xi$ parameter space. First, LHC experiments constrain the product $|\xi m| < 1.5 \times 10^{17}~ \mathrm{GeV}$, for which the excluded parameter space is shown by the purple shaded region in Fig.~\ref{fig:constraint}. Second, torsion balance experiments measuring the strength of a fifth force impose a lower bound on the scalaron mass, indicated by the blue dashed line. Third, the non-observation of an excess photon flux by the INTEGRAL/SPI telescope constrains the scalaron decay width into a pair of photons (Fig.~\ref{fig:decay}), excluding scalaron masses above $m \simeq 0.7~\mathrm{MeV}$, as shown by the green shaded region in Fig.~\ref{fig:constraint}. Taking these bounds together, the allowed parameter space for two of the free parameters of the  Higgs-scalaron mixed model, $m$ and $\xi$, is presented in Fig.~\ref{fig:constraint} for allowed choices of $\lambda$. 

For scenario-I, the observed dark matter relic density is satisfied in two distinct regions, namely - $3\xi m^2 \ll \lambda v^2 $ and $3\xi m^2 \gg\lambda v^2$.
The first corresponds to a fixed scalaron mass $m = 3.6~\mathrm{meV}$ for a wide range of non-minimal couplings, $0 \leq \xi \leq 10^{25}$.
The second allows scalaron masses starting from $m \simeq 10~\mathrm{meV}$ and extending to arbitrarily large values, while $\xi$ is fixed by the relic density, i.e., $ \xi = 1.96\times 10^{26}$, as in Eq.\ref{eq:xi}. 
These two cases are shown in Fig.~\ref{fig:constraint} by the red solid horizontal and vertical contours of constant relic density in the $m$--$\xi$ plane. In the regime, $3\xi m^2 \ll \lambda v^2$, scenario-I marginally survives the fifth-force constraint, as indicated by the blue shaded region.
However, for $3\xi m^2 \gg \lambda v^2$, the apparent absence of an upper bound on the scalaron mass is removed by the LHC constraint, shown by the purple dashed line. As a result, the scalaron mass varies in the range from 10 meV to $770$ meV, while satisfying both relic density and LHC constraints.
This upper bound is stronger by a factor of $\mathcal{O}(10^6)$ compared to the limit derived from the photon flux measured by the INTEGRAL/SPI telescope. In contrast, for $3\xi m^2 \ll \lambda v^2$, the scalaron mass is constrained to be $3.6$ meV, rendering the indirect detection bound irrelevant for both cases within scenario-I.
\begin{figure}[htb!]
 \centering
   \includegraphics[scale=0.6]{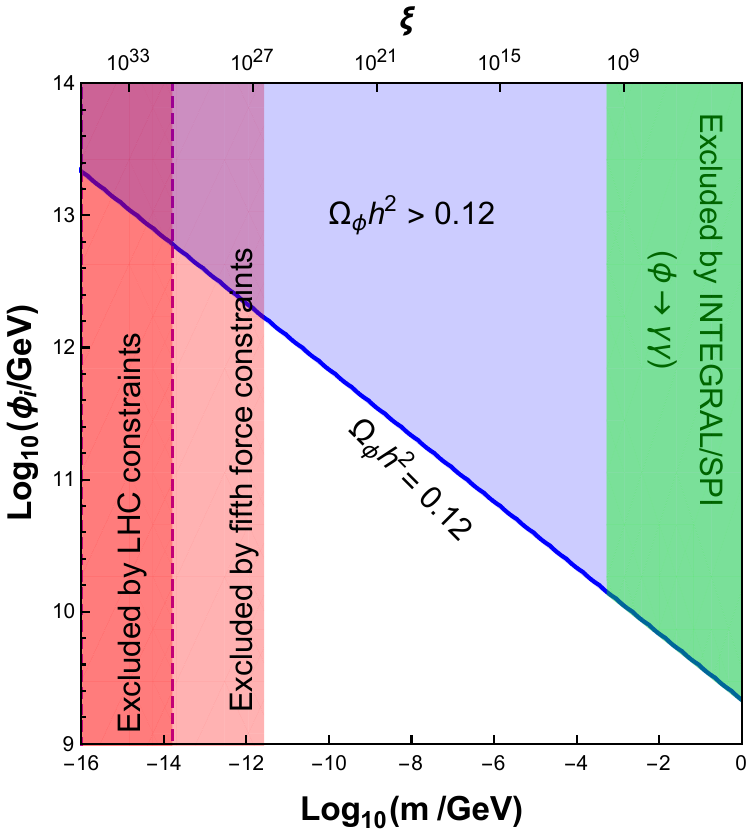}
    \caption{\textit{Scalaron as dark matter in scenario-II} : Constraints on the scalaron mass and its initial field value from the observed DM relic density, fifth force constraint, Higgs mass measurement and INTEGRAL/SPI limits on the excess photon flux from the scalaron decay into a pair of photons.}
    \label{fig:fig3}
\end{figure}

The situation is markedly different for scenario-II. In this case, the INTEGRAL/SPI bound becomes important and determines the upper limit on the scalaron mass, as illustrated by the green shaded region of Fig.~\ref{fig:fig3}, while the lower limit is set by the fifth-force constraint, shown by red shaded region in Fig.~\ref{fig:fig3}. Scenario-II also remains consistent with the LHC constraint, as evident from the contour defined by $3\xi m^2 = \lambda v^2$, shown by the thin gray patch (due to the narrow allowed range of $\lambda$) in Fig.~\ref{fig:constraint}. For each point on the line, the relic density is satisfied for a given $\phi_i$. This is depicted in Fig.\ref{fig:fig3} by a constant contour (blue solid line) of $\Omega_{DM} h^2 = 0.12$. The region above the blue solid line is excluded, as the scalaron would over-saturate the DM density, while the lower region corresponds to the scalaron contributing only a fraction of the DM abundance. We note that the parameter space of the scalaron mass is substantially expanded in scenario-II with the introduction of $\xi$.
 To note, the LHC bound becomes insignificant, unlike scenario-I. This can be seen explicitly by solving $3\xi m^2 = \lambda v^2$ together with $|\xi m| = 1.5 \times 10^{17}~\mathrm{GeV}$, corresponding to the intersection of the black dot-dashed and purple dashed lines in Fig.\ref{fig:constraint}. This yields $m = 1.6 \times 10^{-14}~\mathrm{GeV}$, which is already excluded by the fifth-force constraint, apparent from the purple dashed line in Fig.\ref{fig:fig3}.

To summarize, the lower bound on the scalaron dark matter mass in both scenarios is robustly set by fifth-force experiments. The upper bound, however, depends primarily on the nature of the scalaron Higgs interaction. If the interaction is dominated by the Higgs non-minimal coupling to gravity, the LHC constraint determines the maximum allowed scalaron mass to be $770$ meV. Otherwise, if the Higgs quartic self-coupling dominates, the scalaron mass becomes finely tuned to approximately $3.6$ meV. Finally, in the absence of the scalaron Higgs interaction, the gamma-ray bound from the INTEGRAL/SPI telescope provides the strongest upper limit of $m\approx 0.7$ MeV.

\section{Outlook}
\label{sec:sec6}
In this work, we have studied the scalaron DM dynamics in presence of scalaron-Higgs trilinear interaction, which consists of Higgs quartic coupling and Higgs non-minimal coupling to gravity. Depending on the relative contributions of these couplings the scalaron dynamics gets modified and its mass is significantly altered to account for observed DM density. In particular, we provide a natural explanation for vanishing scalaron-Higgs interaction, leading to DM candidates with masses within meV$-$MeV. 

We note that the interplay between Higgs quartic coupling and the non-minimal coupling in the light DM context is demonstrated with a possibility of detection in highly sensitive X-ray/gamma ray telescopes.
However, there is significant uncertainty over the experimental determination of these couplings, which can affect scalaron dynamics in the early universe. In fact, measuring the Higgs quartic coupling at colliders is very challenging, as it requires cross-section measurements of processes involving the three-Higgs vertex, which are highly suppressed \cite{Fuks:2017zkg,Stylianou:2023tgg}. On the other hand, the limits on the Higgs non-minimal coupling in the low energy effective theory with $R^2$ modification to the Einstein's GR is an unexplored topic, to the best our knowledge. Here, for the first time, we have derived an upper bound on $|\xi m|$ using the Higgs mass and signal strength measurements at the LHC.

Within our analysis of scenario-I, the scalaron dynamics leading to a cold DM candidate at late times is completely determined by the Higgs EWPT. We find that the Higgs non-minimal coupling can take enormously large values in contrast to Higgs-inflation paradigm\cite{Bezrukov:2007ep, Giudice:2010ka,Rubio:2018ogq}, since in the mixed model the effective scale is determined by $|\xi m|$. As long as $m$ is sufficiently small, perturbativity and the unitarity are preserved.

In contrast, within the standard misalignment framework of Scenario~II, the relic density constraint is satisfied for scalaron masses ranging from \(\mathcal{O}(\text{meV})\) to \(\mathcal{O}(\text{MeV})\). In this regime, the scalaron dynamics can be effectively treated as decoupled from the Higgs sector for all practical purposes, with the relic abundance being primarily determined by a sufficiently large initial misalignment. As a result, the details of the electroweak crossover and the subsequent evolution of the Higgs field have a negligible impact on the scalaron evolution. This allows us to safely neglect the precise nature of the electroweak crossover in the analysis of scenario-II.

\section*{Acknowledgment}
The authors thank the anonymous referee for critical comments which led to significant improvement of the manuscript. SGC acknowledges funding from the Anusandhan National Research Foundation (ANRF), Govt. of India, under the National Post-Doctoral Fellowship (File no. PDF/2023/002066). SGC thanks Ananda Dasgupta, Abdul Rahaman Shaikh and Kunal Pandey for useful discussions. DG thanks Sohini Pal for fruitful discussions regarding the LHC constraints. DG acknowledges the Post-doctoral Fellowships from IISER Kolkata and from SINP Kolkata during the course of this work. 

\begin{figure}[h]
    \centering
    \includegraphics[width=0.7\linewidth]{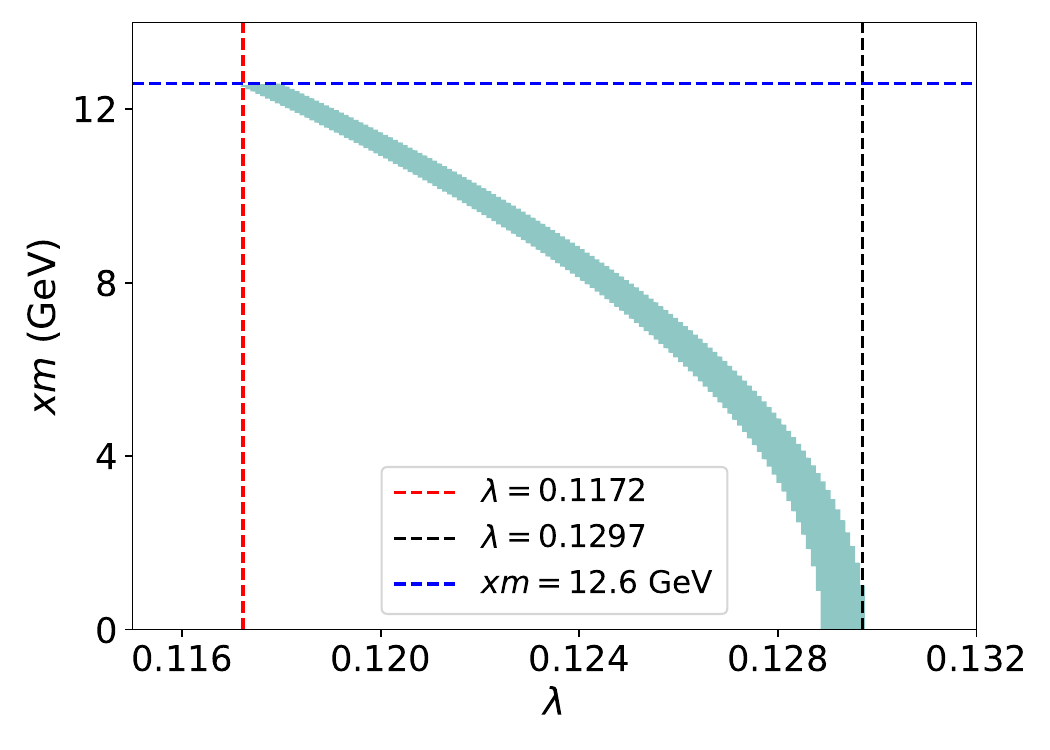}
    \caption{Allowed parameter space of $\lambda$ and $|x m|$ at $95\%$ C.L. using both Higgs mass and signal strength measurements.}
    \label{fig:placeholder}
\end{figure}

\appendix
\section{Derivation of the upper bound on $|\xi m|$}
\label{app}
As noted in Sec.~\ref{sec:sec4lhc}, the measured Higgs mass and the global signal strength are functions of two parameters, $\lambda$ and $xm\equiv\xi v m/M_P$. By demanding that the model simultaneously satisfies the measurements for both the Higgs mass and the global signal strength, we can successfully break the parameter degeneracy. Applying the $95\%$ C.L. allowed ranges for these experimental constraints ($M_H = 125.11 \pm 0.11$ GeV, alongside $\mu = 1.05 \pm 0.06$ from ATLAS and $\mu = 1.002 \pm 0.057$ from CMS), we scanned the parameter space for $\lambda \in [0.01, 1]$ and $|xm| \in (0, 50)$ GeV. The resulting allowed parameter space is shown in Fig.~\ref{fig:placeholder}. 

This combined analysis yields a robust upper bound of $xm \lesssim 12.6$ GeV which translates to : $|\xi m| \lesssim 1.5\times 10^{17}$ GeV, while restricting the self-coupling to the narrow range of $\lambda \approx 0.117 - 0.13$. It is evident that all of the cases, i.e., $3\xi m^2\ll \lambda v^2$, $3\xi m^2\gg \lambda v^2$ and $3\xi m^2\approx \lambda v^2$ can be realized for  $\lambda$ and $\xi m$ values allowed by the LHC constraints.

\bibliography{ref}
\bibliographystyle{JHEP}
\end{document}